# A discrepancy of $10^7$ in experimental and theoretical density detection limits of aerosol particles by surface nonlinear light scattering

Arianna Marchioro[1], Thaddeus W. Golbek[2], Adam S. Chatterley[2], Tobias Weidner[2], and Sylvie Roke[1]*

[1]Laboratory for Fundamental BioPhotonics (LBP), Institute of Bioengineering (IBI), and Institute of Materials Science (IMX), School of Engineering (STI), and Lausanne Centre for Ultrafast Science (LACUS), Ecole Polytechnique Fédérale de Lausanne (EPFL), CH-1015 Lausanne, Switzerland.
[2]Department of Chemistry, Aarhus University, 8000 Aarhus C, Denmark; *e-mail: sylvie.roke@epfl.ch

ARISING FROM In situ analysis of the bulk and surface chemical compositions of organic aerosol particles: Qian et al. Communications Chemistry https://doi.org/10.1038/s42004-022-00674-8 (2022)

**Abstract**
Understanding the interfacial properties of aerosol particles is important for science and medicine, crucial for air quality, human health, and environmental chemistry. Qian et al. presented vibrational sum frequency scattering (SFS) measurements of organic molecules on aerosol particles. Relating an aerosol sample with a 40 nm average size and 10 - 300 nm particle size distribution at a density of a $10^6$ particles / mL to vibrational sum frequency scattering spectra recorded in a different apparatus, it was concluded that the vibrational spectra reported on the surface structure of the particles in the aerosol. Here, we show that the SF scattering power of such small particles with a density of a $10^6$ particles / mL is ~$10^7$ smaller than the detection limit of the presented SFS experiment. We determine the detectable number density of particles, both theoretically and experimentally, to demonstrate the limits of the SFS method. We also propose possible reasons for the $10^7$ order of magnitude discrepancy.

**Introduction**

Aerosol science would benefit greatly from understanding the interfacial structure of airborne particles. However, measuring the surface structure *in situ* is challenging because of the weak second-order nonlinearity needed to generate a surface response, the dispersive / scattering nature of the sample, the dilution of particles (~ $10^6$ / mL), and the size range of the majority of them (10 - 300 nm, in diameter).[1] Indeed, vibrational surface SFS spectra as well as non-resonant surface second harmonic scattering (SHS, Fig. 1A, which is comparable) measured from particles dispersed in solution of this size range have been reported with samples containing ~$10^{11}$ particles / mL.[2,3] There is no principle difference in the mechanism and rules that govern nonlinear scattering for particles that are airborne or dispersed in solution,[4] with the difference in surrounding medium being incorporated by the difference in refractive index between the particle and surrounding medium. Since the 1980's numerous groups have worked on developing and verifying nonlinear light scattering theories, and these theories, like their linear counterpart, generally agree with nonlinear light scattering experiments.[5] Indeed, the single difference between linear and nonlinear light scattering is the induced polarization of the material that is used as a source term. Therefore, the results by Qian et al.[1] raise questions as to the nature of the source of the SF scattering object(s), as there is a difference on the order of ~ $10^7$ between the expected scattering power and the detected spectra.

Here, we first explicitly determine the number of particles that can be reasonably measured for a certain size in SHS/SFS experiments. To do so, we combine theoretical predictions per particle with experiment. We will first consider the throughput of the SFS/SHS experiment and consider its size and number density dependence, then provide a signal-to-noise ratio analysis of various comparable experiments. Using this analysis in combination with SFS and SHS experiments performed on the same samples, we determine the detection limit in terms of particle density for a certain size. Finally, we discuss several explanations for the difference in terms of detected intensity and expected aerosol intensity.

**Size dependence.** Vibrational SFS and non-resonant SHS are second-order nonlinear optical techniques. The intensity of the generated photons obeys the following expression:

$$I(\omega_0) \propto N_p |\Gamma^{(2)}(R, \theta, \chi^{(2)})|^2 \frac{E_1 E_2}{\tau A} f \qquad (1)$$

with $N_p$ the particle density, $\Gamma^{(2)}(R, \theta, \chi^{(2)})$ the effective (single) particle susceptibility, which depends on the radius $R$, the scattering angle $\theta$, the (surface) susceptibility of the particle $\chi^{(2)}$, $E_i$ the pulse energies of the participating beams, $\tau$ the pulse duration, $A$ the overlap area and $f$ the repetition rate. The effective particle susceptibility that determines the response of a single particle $\Gamma^{(2)}(R, \theta, \chi^{(2)})$ is determined by the single particle light-matter interaction process and is highly size dependent. $\Gamma^{(2)}$ also contains effects of absorption/linear scattering as it is also a function of the electromagnetic field functions.[6] For $R \sim < 200$ nm, $\Gamma^{(2)} \propto R^3$ ($I(\omega_0) \propto R^6$), and for larger particles, this size dependence levels off, reaching $I(\omega_0) \propto R^3$ at $R \sim 1000$ nm. For $R < 200$ nm, the Rayleigh-Gans-Debye approximation, one of the approximate solutions to the Maxwell equations, works well to describe the data. For larger sizes, it becomes more approximate, and depending on the conditions nonlinear Mie theory needs to be used.[4] Nonlinear Mie theory offers an exact solution for both linear and nonlinear scattering, assuming the single scattering particles are spherical. Dipolar and quadrupolar scattering both lead to identical size-dependent behaviors.[4,7] The scattering pattern is also highly size dependent with scattering maxima appearing between 90°-55° for $R$ < 50 nm, which gradually move forward. For water droplets in air, however, the refractive index contrast ensures that for micron-sized or larger particles scattering light is emitted in every direction.[4] Particles in air have a bigger linear refractive index difference between the bulk and the particle medium compared to solid/liquid dispersions, which generally have a smaller refractive index contrast and therefore exhibit less losses due to linear scattering of the incident beams. We can expect that the estimations based on the systems used in this work are over-estimating the actual scattering efficiencies.

**Signal to Noise Ratio.** To relate the experimental throughput of the experiments conducted by Rao and co-workers[1] we first compare their signal to noise ratio (SNR) to previously published vibrational SFS and non-resonant SHS data, and then perform vibrational SFS and non-resonant SHS measurements on hexadecane nanodroplets. Table 1 shows the experimental parameters that relate the 3 published experiments, as well as the parameters used to collect the data in Figs. 1C and 1D. The SFS experiment reported by Rao and co-workers[1] has a comparable SNR compared to previously published SFS data and a 1.4 - 2.4 smaller SNR to non-resonant SHS. Because SHS and SFS have comparable SNRs, we retrieve the particle vs size dependence from SHS. The primary reason to do so is that the SHS intensity of any particle surface can be compared both experimentally and theoretically to the known incoherent bulk response of neat water, which therefore represents a calibration benchmark. This incoherent SH light is known as hyper Rayleigh scattering, and is subtracted from the raw data, to obtain the pure particle scattering (Eq. S1). With a known hyperpolarizability tensor of water, it is therefore possible to explicitly compute the theoretical response that matches 10 % of the magnitude of the intensity recorded with SNR = 1. We take this as the detection limit of the SFS and non-resonant SHS experiments of particles. Thus, we compute using the theory in Refs.[8,9], as a function of particle radius, which number density of particles is needed to generate an intensity that matches 10 % of the incoherent neat bulk water intensity in the SSS polarization combination (all beams polarized perpendicular to the scattering plane). We insert typical values for the surface susceptibility ($\chi^{(2)}$ =$10^{-22}$ m²/V) and the surface potential ($\Phi_0$ = 100 mV), which represent the surface properties of the particles.

**Table 1: Experimental parameters.** PPP (SSS) refers to all beams polarized in (perpendicular to) the scattering plane.

|  | SFS as in Ref [1] | SFS as in Refs.[2,13] | SHS as in Refs.[11,12] | SFS, Fig. 1C | SHS Fig. 1D |
|---|---|---|---|---|---|
| Fundamental wavelength | 1025 nm | 800 nm | 1030 nm | 517 nm | 1032 nm |
| Repetition rate | 100 kHz | 1 kHz | 200 kHz | 10 kHz | 200 kHz |
| OPA range | 2500 – 4500 nm | 2600 – 20000 nm | NA | 2300 – 15000 nm | NA |
| IR pulse energy at the sample ($\mu$J) | 2 | 5-10 | 0.3 | 13.6 | 0.4 |
| Beam waist (diameter, $\mu$m) | 80 | 340 | 110 | 400 | 110 |
| Fluence (mJ/cm$^2$) | 39.8 | 5.5 | 3.2 | 10.8 | 4.2 |
| VIS pulse linewidth (cm$^{-1}$) | 8 cm$^{-1}$ | 12 cm$^{-1}$ | NA | 0.18 nm | NA |
| VIS pulse energy ($\mu$J) | 6 | 5-10 | NA | 7.5 | NA |
| Collection angle range ($\theta$) | 90° $\pm$ 30° | $\theta_{max} \pm 10°$ | -90° < $\theta$ < +90° (3.4° / $\theta$) | $\theta_{max} \pm 10°$ | -90° < $\theta$ < +90° (approx. 4° / $\theta$) |
| Max. SNR ratio (PPP or SSP) | 11 | 5 - 15 | 16 - 26 | 36 | 15 |
| SNR ratio SSS bulk H$_2$O |  |  | 18 - 21 |  | ~ 15 |

**Estimation of particle density vs size dependence.** The line in Fig. 1B shows the detection limit in terms of # particles / mL as a function of the radius of the particles, based on extrapolating the various size-dependent models.[4] For a 20 nm radius, this means 2.5 x 10$^{13}$ particles / mL are needed to generate the desired SF intensity. Moving to larger particles, we arrive at 5.6 x 10$^{10}$ particles / mL (50 nm radius), 4.2 x 10$^7$ (500 nm radius), and finally, extrapolating to non-resonant SH imaging, 1 for objects in the size range of 5 – 10 microns.[10] For diameters close to 100, 200 and 300 nm, data points from angle-resolved non-resonant second harmonic scattering measurements from the surface of silica particles dispersed in water[11,12] are shown. Based on this result, we estimate that 10$^6$ particles / mL can only generate a detectable response if R > 1000 nm. Alternatively, 40 nm particles would have to be present at a density of 2.5 x 10$^{13}$ particles / mL, as indicated by the dashed arrow. Neither of these criteria are met in Ref.[1]

Fig. 1C shows SFS spectra of deuterated hexadecane droplets with a 109 ± 1 nm average hydrodynamic radius stabilized with 8 mM sodium dodecyl sulfate (SDS), following the protocol of Chen et al.[13] and summarized in the Supplementary Methods. SFS spectra were recorded with particle densities between 10$^{11}$ / mL and 10$^{13}$ / mL, achieved by diluting a stock emulsion (size distribution is shown in the inset). In agreement with Fig. 1B, the recorded SF intensity vanishes below a particle density of 10$^{12}$ / mL. Fig. 1D shows data measured by non-resonant SHS of the same sample diluted with 0.8 mM SDS solution. In both experiments, no detectable signal can be obtained below 4.9 x 10$^9$ particles / mL, in agreement with predictions of Fig. 1B for identically-sized objects.

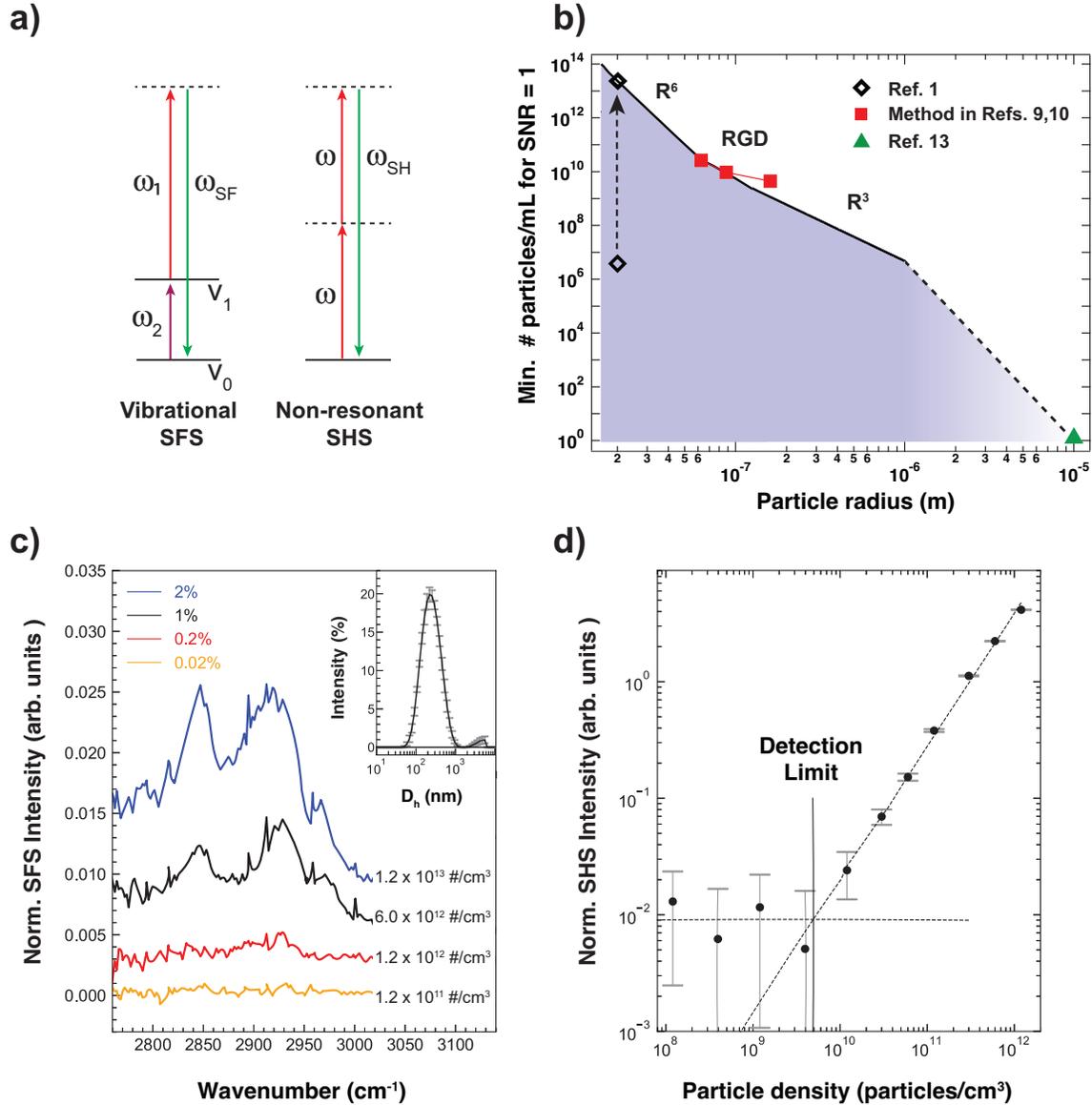

**Figure 1: Number density vs size dependence**. **a)**: Illustration of energy schemes of vibrational sum frequency scattering and non-resonant second harmonic scattering. **b)**: Number density necessary to generate a signal-to-noise ratio = 1 as a function of particle size, assuming the minimal detectable intensity comprises 10 % of the non-resonant SHS response of neat bulk water. Black diamonds, red squares and green triangle correspond to data points extracted from the indicated references. The colored area is inaccessible to current non-resonant SHS and vibrational SFS instruments. **c), d)**: Vibrational SFS (**c**) and non-resonant SHS measurements (**d**) of SDS stabilized $d_{34}$-hexadecane nanodroplets in $D_2O$ measured at different droplet densities, ranging from $1.2 \times 10^{13}$ /cm$^3$ (blue trace) to $1.2 \times 10^{11}$ /cm$^3$ (orange trace, which we consider the detection limit). The number of droplets is computed from the used volume in combination with the hydrodynamic diameter ($D_h$) measured by DLS (shown in the inset of **c**). The spectra in the SFS plot are offset for clarity. The SHS measured intensity (**d**, black dots) and standard error (grey error bars) are plotted with trend lines for both the intensity above and below the noise level (dashed lines). The detection limit was determined to be at the intersection of both lines at $4.9 \times 10^9$ /cm$^3$.

**Discussion and Conclusions**. Although the values of $\chi^{(2)}$, the scattering angle range, the precise value of the refractive indices, or the surface potential may change, given the difference examined here (a factor of $10^7$ for a particle of 40 nm in diameter) these details are relatively unimportant as they would create a maximum difference in the scattering intensity in the range 0.1 – 10. Since the instrument used by Rao and co-workers[1] is not outperforming the instruments used here (Table 1), this brings about the question what actually generates the unexpectedly strong SFS response. Based on nonlinear light scattering theory summarized by Eq. (1), there are only two candidates for increasing the scattered intensity by x $10^7$: The size of the particles, with the scattering objects being much larger than reported, or $\chi^{(2)}$, which should then increase by ~$10^3$. A single particle of a few microns in size could easily overpower the emission of all other particles and generate the measured SF intensity, for example $\frac{I(1\ \mu m)}{I(10\ nm)} \sim (10^2)^3$. To get insight here, one would have to measure the SF scattering patterns, as was done in Ref.[2] A time-dependent intensity trace would provide information about the stability of the intensity.[14] The second option might revolve around the crystallization of surfactant in a semi-crystalline layer on the surface of the particles. Extremely dilute crystallites have also shown to produce large SFS intensities.[15] It is additionally known in aerosol literature that due to differences in evaporation rates, particles size, morphology and crystallinity can drastically change once they are released in the aerosol chamber.[16–18]

Summarizing, performing vibrational surface SFS from aerosols with "an average size of 40 nm and a density of $10^6$ particles / mL"[1] is at odds with understanding based upon nonlinear light scattering theory as well as state of the art experiments conducted by numerous labs, on (metal/plasmonic) particles, droplets, harmonophores, liposomes and other objects with vastly different compositions in both liquid and solid media. The nonlinear light scattering data and theory obtained since the 1980's generally agrees well with one another, and in that sense shows the same type of confidence as one has with linear light scattering.[19] That Ref.[1] shows data that deviates by a factor of ~$10^7$ from current state of the art data and understanding is therefore extremely puzzling to us.

**Acknowledgements**

A.M and S.R thank the Julia Jacobi Foundation. A.M thanks the Swiss National Science Foundation (Ambizione grant number PZ00P2_174146). T.W.G, A.S.C and T.W. acknowledge funding from the Novo Nordisk Foundation (Facility Grant NanoScat, No. NNF18OC0032628). T.W.G thanks the Lundbeck Foundation for a postdoc fellowship (postdoc grant R322-2019-2461).

**Author contributions**

S.R. conceived and designed the work. A.M carried out the theoretical calculations. T.W.G and A.S.C performed the SFS/SHS experiments. A.M, T.W.G and S.R wrote the manuscript. T.W and S.R. directed the work. All authors discussed the results and contributed to the manuscript.

**Data availability**

The data sets generated and analyzed during the current study are available from the corresponding author upon reasonable request.

**Competing interests**

The authors declare no competing interests.

**References**

1. Qian, Y. *et al.* In situ analysis of the bulk and surface chemical compositions of organic aerosol particles. *Comm. Chem.* **5**, 58 (2022).

2. Roke, S. *et al.* Vibrational Sum Frequency Scattering from a Submicron Suspension. *Phys. Rev. Lett.* **91**, 1–4 (2003).

3. Schönfeldová, T. *et al.* Lipid Melting Transitions Involve Structural Redistribution of Interfacial Water. *J. Phys. Chem. B* **125**, 12457–12465 (2021).

4. Beer, A. G. F. de & Roke, S. Nonlinear Mie theory for second-harmonic and sum-frequency scattering. *Physical Review B* **79**, 155420–1--155420--9 (2009).

5. Roke, S. & Gonella, G. Nonlinear Light Scattering and Spectroscopy of Particles and Droplets in Liquids. *Annu. Rev. Phys. Chem.* **63**, 353–378 (2012).

6. Kulik, S., Pullanchery, S. & Roke, S. Vibrational Sum Frequency Scattering in Absorptive Media: A Theoretical Case Study of Nano-objects in Water. *J. Phys. Chem. C* **124**, 23078–23085 (2022).

7. Held, H., Lvovsky, A. I., Wei, X. & Shen, Y. R. Bulk contribution from isotropic media in surface sum-frequency generation. *Phys. Rev. B* **66**, 205110 (2002).

8. Gonella, G., Lütgebaucks, C., Beer, A. G. F. de & Roke, S. Second Harmonic and Sum-Frequency Generation from Aqueous Interfaces Is Modulated by Interference. *J. Phys. Chem. C* **120**, 9165–9173 (2016).

9. Lütgebaucks, C., Gonella, G. & Roke, S. Optical label-free and model-free probe of the surface potential of nanoscale and microscopic objects in aqueous solution. *Phys. Rev. B* **94**, 195410 (2016).

10. Macias-Romero, C., Nahalka, I., Okur, H. I. & Roke, S. Optical Imaging of Surface Chemistry and Dynamics in Confinement. *Science* **357**, 784–788 (2017).

11. Marchioro, A. *et al.* Surface Characterization of Colloidal Silica Nanoparticles by Second Harmonic Scattering: Quantifying the Surface Potential and Interfacial Water Order. *J. Phys. Chem. C* **123**, 20393–20404 (2019).

12. Bischoff, M., Biriukov, D., Předota, M., Roke, S. & Marchioro, A. Surface Potential and Interfacial Water Order at the Amorphous $TiO_2$ Nanoparticle/Aqueous Interface. *J. Phys. Chem. C* **124**, 10961–10974 (2020).

13. Chen, Y., Jena, K. C., Lütgebaucks, C., Okur, H. I. & Roke, S. Three Dimensional Nano "Langmuir Trough" for Lipid Studies. *Nano Lett.* **15**, 5558–5563 (2015).

14. Dadap, J. I., Aguiar, H. B. de & Roke, S. Nonlinear light scattering from clusters and single particles. *J. Chem. Phys.* **130**, 214710 (2009).

15. Aguiar, H. B. de, Beer, A. G. F. de & Roke, S. The Presence of Ultralow Densities of Nanocrystallites in Amorphous Poly(lactic acid) Microspheres. *J. Phys. Chem. B* **117**, 8906–8910 (2013).

16. Gregson, F. K. A. *et al.* Studies of competing evaporation rates of multiple volatile components from a single binary-component aerosol droplet. *Phys. Chem. Chem. Phys.* **21**, 9709–9719 (2019).

17. Gregson, F. K. A., Robinson, J. F., Miles, R. E. H., Royall, C. P. & Reid, J. P. Drying Kinetics of Salt Solution Droplets: Water Evaporation Rates and Crystallization. *J. Phys. Chem. B* **123**, 266–276 (2019).

18. Ingram, S. *et al.* Accurate Prediction of Organic Aerosol Evaporation Using Kinetic Multilayer Modeling and the Stokes–Einstein Equation. *J. Phys. Chem. A* **125**, 3444–3456 (2021).

19. Hulst, H. C. van de. *Light Scattering by Small Particles*, Dover Publications Inc., New York, 1981.